\documentstyle[aps]{revtex}
\newcommand{\bel}[1]{\begin{equation}\label{#1}}
\newcommand{\be}{\begin{equation}}
\newcommand{\ee}{\end{equation}}

\newcommand{\beal}[1]{\begin{eqnarray}\label{#1}}
\newcommand{\bea}{\begin{eqnarray}}
\newcommand{\eea}{\end{eqnarray}}

\newcommand{\bean}{\begin{eqnarray*}}
\newcommand{\eean}{\end{eqnarray*}}

\newcommand{\ba}{\begin{array}}
\newcommand{\ea}{\end{array}}

\newcommand{\bab}{\begin{abstract}}
\newcommand{\eab}{\end{abstract}}

\newcommand{\bml}{\begin{mathletters}}
\newcommand{\eml}{\end{mathletters}}

\newcommand{\q}{\quad}
\newcommand{\qq}{\quad\quad}

\newcommand{\bfm}[1]{\mbox{\boldmath $#1$}}

\newcommand{\dv}{\partial}

\newcommand{\bam}{\left( \begin{array}}
\newcommand{\eam}{\end{array} \right)}

\newcommand{\bamq}[4]{\left( \begin{array}{cccc}{#1}&{#2}&{#3}&{#4}\\}
\newcommand{\bamc}[5]{\left( \begin{array}{ccccc}{#1}&{#2}&{#3}&{#4}&{#5}\\}

\newcommand{\lrw}{\leftrightarrow}

\newcommand{\cg}{\gamma}
\newcommand{\dg}{\delta}
\newcommand{\eg}{\epsilon}

\newcommand{\vpg}{\varphi}

\def\xc{{\mathchoice {\setbox0=\hbox{$\displaystyle\rm C$}\hbox{\hbox
to0pt{\kern0.4\wd0\vrule height0.9\ht0\hss}\box0}}
{\setbox0=\hbox{$\textstyle\rm C$}\hbox{\hbox
to0pt{\kern0.4\wd0\vrule height0.9\ht0\hss}\box0}}
{\setbox0=\hbox{$\scriptstyle\rm C$}\hbox{\hbox
to0pt{\kern0.4\wd0\vrule height0.9\ht0\hss}\box0}}
{\setbox0=\hbox{$\scriptscriptstyle\rm C$}\hbox{\hbox
to0pt{\kern0.4\wd0\vrule height0.9\ht0\hss}\box0}}}}

\def\xg{{\mathchoice {\setbox0=\hbox{$\displaystyle\rm G$}\hbox{\hbox
to0pt{\kern0.4\wd0\vrule height0.9\ht0\hss}\box0}}
{\setbox0=\hbox{$\textstyle\rm G$}\hbox{\hbox
to0pt{\kern0.4\wd0\vrule height0.9\ht0\hss}\box0}}
{\setbox0=\hbox{$\scriptstyle\rm G$}\hbox{\hbox
to0pt{\kern0.4\wd0\vrule height0.9\ht0\hss}\box0}}
{\setbox0=\hbox{$\scriptscriptstyle\rm G$}\hbox{\hbox
to0pt{\kern0.4\wd0\vrule height0.9\ht0\hss}\box0}}}}

\def\xi{{\rm I\!I}}

\def\xo{{\mathchoice {\setbox0=\hbox{$\displaystyle\rm O$}\hbox{\hbox
to0pt{\kern0.4\wd0\vrule height0.9\ht0\hss}\box0}}
{\setbox0=\hbox{$\textstyle\rm O$}\hbox{\hbox
to0pt{\kern0.4\wd0\vrule height0.9\ht0\hss}\box0}}
{\setbox0=\hbox{$\scriptstyle\rm O$}\hbox{\hbox
to0pt{\kern0.4\wd0\vrule height0.9\ht0\hss}\box0}}
{\setbox0=\hbox{$\scriptscriptstyle\rm O$}\hbox{\hbox
to0pt{\kern0.4\wd0\vrule height0.9\ht0\hss}\box0}}}}

\def\xq{{\mathchoice {\setbox0=\hbox{$\displaystyle\rm
Q$}\hbox{\raise
0.15\ht0\hbox to0pt{\kern0.4\wd0\vrule height0.8\ht0\hss}\box0}}
{\setbox0=\hbox{$\textstyle\rm Q$}\hbox{\raise
0.15\ht0\hbox to0pt{\kern0.4\wd0\vrule height0.8\ht0\hss}\box0}}
{\setbox0=\hbox{$\scriptstyle\rm Q$}\hbox{\raise
0.15\ht0\hbox to0pt{\kern0.4\wd0\vrule height0.7\ht0\hss}\box0}}
{\setbox0=\hbox{$\scriptscriptstyle\rm Q$}\hbox{\raise
0.15\ht0\hbox to0pt{\kern0.4\wd0\vrule height0.7\ht0\hss}\box0}}}}

\def\xs{{\mathchoice
{\setbox0=\hbox{$\displaystyle     \rm S$}\hbox{\raise0.5\ht0\hbox
to0pt{\kern0.35\wd0\vrule height0.45\ht0\hss}\hbox
to0pt{\kern0.55\wd0\vrule height0.5\ht0\hss}\box0}}
{\setbox0=\hbox{$\textstyle        \rm S$}\hbox{\raise0.5\ht0\hbox
to0pt{\kern0.35\wd0\vrule height0.45\ht0\hss}\hbox
to0pt{\kern0.55\wd0\vrule height0.5\ht0\hss}\box0}}
{\setbox0=\hbox{$\scriptstyle      \rm S$}\hbox{\raise0.5\ht0\hbox
to0pt{\kern0.35\wd0\vrule height0.45\ht0\hss}\raise0.05\ht0\hbox
to0pt{\kern0.5\wd0\vrule height0.45\ht0\hss}\box0}}
{\setbox0=\hbox{$\scriptscriptstyle\rm S$}\hbox{\raise0.5\ht0\hbox
to0pt{\kern0.4\wd0\vrule height0.45\ht0\hss}\raise0.05\ht0\hbox
to0pt{\kern0.55\wd0\vrule height0.45\ht0\hss}\box0}}}}

\def\xt{{\mathchoice {\setbox0=\hbox{$\displaystyle\rm
T$}\hbox{\hbox to0pt{\kern0.3\wd0\vrule height0.9\ht0\hss}\box0}}
{\setbox0=\hbox{$\textstyle\rm T$}\hbox{\hbox
to0pt{\kern0.3\wd0\vrule height0.9\ht0\hss}\box0}}
{\setbox0=\hbox{$\scriptstyle\rm T$}\hbox{\hbox
to0pt{\kern0.3\wd0\vrule height0.9\ht0\hss}\box0}}
{\setbox0=\hbox{$\scriptscriptstyle\rm T$}\hbox{\hbox
to0pt{\kern0.3\wd0\vrule height0.9\ht0\hss}\box0}}}}

\def\xz{{\mathchoice {\hbox{$\sf\textstyle Z\kern-0.4em Z$}}
{\hbox{$\sf\textstyle Z\kern-0.4em Z$}}
{\hbox{$\sf\scriptstyle Z\kern-0.3em Z$}}
{\hbox{$\sf\scriptscriptstyle Z\kern-0.2em Z$}}}}

\newcommand{\fs}{\footnotesize}

\newcommand{\bii}{\begin{itemize}}
\newcommand{\eii}{\end{itemize}}

\newcommand{\ben}{\begin{enumerate}}
\newcommand{\een}{\end{enumerate}}

\newcommand{\bq}{\begin{quote}}
\newcommand{\eq}{\end{quote}}

\newcommand{\bc}{\begin{center}}
\newcommand{\ec}{\end{center}}

\newcommand{\btb}{\begin{table}}
\newcommand{\etb}{\end{table}}

\newcommand{\bt}{\begin{tabular}}
\newcommand{\et}{\end{tabular}}

\newcommand{\br}{\begin{flushright}}
\newcommand{\er}{\end{flushright}}

\newcommand{\bl}{\begin{flushleft}}
\newcommand{\el}{\end{flushleft}}

\newcommand{\vs}[1]{\vspace*{#1}}

\newcommand{\new}{\pagebreak}

\newcommand{\bref}{\begin{references}}
\newcommand{\eref}{\end{references}}

\newcommand{\bb}{\begin{thebibliography}{99}}
\newcommand{\eb}{\end{thebibliography}}
\newcommand{\bi}{\bibitem}

\newcommand{\btp}{\begin{titlepage}}
\newcommand{\etp}{\end{titlepage}}

\newcommand{\go}{\section{Introduction}}
\newcommand{\con}{\section{Conclusions}}

\newcommand{\axp}[3]{Ann.~Phys.~(NY)                    {\bf #1},  #2  (19#3)}

\newcommand{\ixa}[3]{Int.~J.~Mod.~Phys.~A               {\bf #1},  #2  (19#3)}

\newcommand{\jxe}[3]{J.~Math.~Phys.                     {\bf #1},  #2  (19#3)} 

\newcommand{\jxg}[3]{J.~Phys.~A                         {\bf #1},  #2  (19#3)}

\newcommand{\mxb}[3]{Mod.~Phys.~Lett.~A                 {\bf #1},  #2  (19#3)}

\newcommand{\nxb}[3]{Nucl.~Phys.                        {\bf #1},  #2  (19#3)}

\newcommand{\nxd}[3]{Nuovo Cimento                      {\bf #1},  #2  (19#3)}

\newcommand{\pxa}[3]{Phys.~Essay.                       {\bf #1},  #2  (19#3)}

\newcommand{\pxf}[3]{Phys.~Rev.~D                       {\bf #1},  #2  (19#3)}

\newcommand{\pxh}[3]{Phys.~Rev.~Lett.                   {\bf #1},  #2  (19#3)}
\newcommand{\pxi}[3]{Phys.~Lett.                        {\bf #1},  #2  (19#3)}

\newcommand{\pxxa}[3]{Prog.~Theor.~Phys.                {\bf #1},  #2  (19#3)}

\newcommand{\xxx}[3]{{\bf #1},  #2  (19#3)}

\newcommand{\co}{\mbox{\boldmath $\cal C$}}
\newcommand{\qu}{\mbox{\boldmath $\cal H$}}
\newcommand{\oct}{\mbox{\boldmath $\cal O$}}
\newcommand{\rea}{\mbox{\boldmath $\cal R$}}
\newcommand{\glr}{GL(8, \rea )}
\newcommand{\glc}{GL(4,  \co )}

\title{OCTONIONIC DIRAC EQUATION}

\author{Stefano De Leo$^{(a,b)}$ and 
Khaled Abdel-Khalek$^{(a)}$} 
\address{$^{(a)}$~Dipartimento di Fisica - Universit\`a di Lecce\\
$^{(b)}$~Istituto Nazionale di Fisica Nucleare, Sezione di Lecce\\
- Lecce, 73100, Italy -}

\date{Revised Version, July 1996}

\draft

\begin{document}

\maketitle

\bab
In order to obtain a consistent formulation of octonionic quantum mechanics 
(OQM), we introduce left-right barred operators. Such operators enable us to 
find the translation rules between octonionic numbers and $8\times 8$ real 
matrices (a translation is also given for $4\times 4$ complex matrices). 
We develop an octonionic relativistic free wave 
equation, linear in the derivatives. Even if the wave functions are only 
one-component we show that four independent solutions, corresponding to 
those of the Dirac equation, exist.         
\eab

\renewcommand{\thefootnote}{\sharp\arabic{footnote}}

\go

From the sixties onwards, there has been renewed and intense interest in 
the use of octonions in physics~\cite{gur1}. The octonionic algebra has been 
in fact linked with a number of interesting subjects: structure 
of interactions~\cite{pais}, $SU(3)$ color symmetry and quark 
confinement~\cite{gur2,mor}, standard model gauge group~\cite{dix}, 
exceptional GUT groups~\cite{gur3}, Dirac-Clifford algebra~\cite{edm}, 
nonassociative Yang-Mills theories~\cite{jos1,jos2}, space-time symmetries in 
ten dimensions~\cite{dav}, supersymmetry and supergravity 
theories~\cite{sup1,sup2}. Moreover, the recent successful application of 
quaternionic numbers in quantum mechanics~\cite{adl,adl1,qua1,qua2,qua3}, 
in particular in formulating a quaternionic Dirac 
equation~\cite{dir1,dir2,dir3,dir4}, suggests 
going one step further and using octonions as underlying numerical field. 

In this work, we overcome the problems due to the nonassociativity of the 
octonionic algebra by introducing left-right barred operators 
(which will be sometimes called barred octonions). Such operators 
complete the mathematical material introduced in the recent papers 
of Joshi {\it et al.}~\cite{jos1,jos2}. Then, we 
investigate their relations  to $\glr$ and $\glc$. Establishing this 
relation we find 
interesting translation rules, which gives us the opportunity to formulate 
a consistent OQM. 

The philosophy behind the translation can be concisely expressed by the 
following sentence: ``There exists at least one version of octonionic
quantum mechanics where the standard quantum mechanics is reproduced''. 
The use of a complex scalar product (complex geometry)~\cite{hor}  
will be the main tool to obtain OQM. 

We wish to stress that translation rules don't imply that our octonionic 
quantum world (with complex geometry) is equivalent to the standard quantum 
world. When translation 
fails the two worlds are not equivalent. An interesting case 
can be supersymmetry~\cite{rk}. 

Similar translation rules, between quaternionic quantum mechanics (QQM)
with complex geometry and standard quantum mechanics, have been recently 
found~\cite{qua2}. As an application, such rules can be exploited in 
reformulating in a natural way the electroweak sector of the standard 
model~\cite{qua3}.

In section II, we discuss octonionic algebra and introduce barred 
operators. Then, in Section III, we investigate 
the relation between barred octonions and $8\times 8$ real matrices. 
In this section, we also give the translation rules between octonionic barred 
operators and $\glc$, which will be very useful in formulating our OQM 
(full details of the mathematical material appear elsewhere~\cite{jmp}).  
In section IV, we explicitly develop an 
octonionic Dirac equation and suggest possible difference between complex 
and octonionic quantum theories. In the final section we draw our conclusions.

\section{Octonionic barred operators}

We can characterize the algebras $\rea$, $\co$, $\qu$ and 
$\oct$ by the concept of {\tt division algebra} (in which one has no nonzero 
divisors of zero). Octonions, which locate a nonassociative division algebra, 
can be represented by seven imaginary units $(e_1 , \ldots ,e_7 )$ 
and $e_0\equiv 1$: 
\be
{\cal O} =r_{0}+\sum_{m=1}^{7} r_{m}e_{m} 
\qq (~r_{0,...,7}~~ \mbox{reals}~) \q . 
\ee 
These seven imaginary units, $e_{m}$, obey the noncommutative
and nonassociative algebra
\be
e_{m}e_{n}=-\dg_{mn}+ \eg_{mnp}e_{p} \qq 
(~\mbox{{\fs $m, \; n, \; p =1,..., 7$}}~) 
\q ,
\ee
with $\eg_{mnp}$ totally antisymmetric and equal to unity for the seven 
combinations 
$123, \; 145, \; 176, \; 246, \; 257, \; 347 \; \mbox{and} \; 365$. 
The norm, $N({\cal O})$, for the octonions is defined by
\be
N({\cal O})=({\cal O}^{\dag}{\cal O})^{\frac{1}{2}}=
({\cal O}{\cal O}^{\dag})^{\frac{1}{2}}=
(r_{0}^{2}+  ... + r_{7}^{2})^{\frac{1}{2}} \q ,
\ee
with the octonionic conjugate $o^{\dag}$ given by
\be
{\cal O}^{\dag}=r_{0}-\sum_{m=1}^{7} r_{m}e_{m} \q . 
\ee 
The inverse is then 
\be
{\cal O}^{-1}={\cal O}^{\dag}/N({\cal O}) \qq (~{\cal O}\neq 0~) \q .
\ee

We can define an {\tt associator} (analogous to the usual algebraic 
commutator) as follows
\bel{ass}
\{x, \; y , \; z\}\equiv (xy)z-x(yz) \q ,
\ee
where, in each term on the right-hand, we must, first of all, perform 
the multiplication in brackets. 
Note that for real, complex and quaternionic numbers the associator is 
trivially null. For octonionic imaginary units we have
\bel{eqass}
\{e_{m}, \; e_{n}, \; e_{p} \}\equiv(e_{m}e_{n})e_{p}-e_{m}(e_{n}e_{p})=
2 \eg_{mnps} e_{s} \q ,
\ee
with $\eg_{mnps}$ totally antisymmetric and equal to unity for the seven 
combinations 
\[ 
1247, \; 1265, \; 2345, \; 2376, \; 3146, \; 3157 \; \mbox{and} \; 4567 \q .
\]
Working with octonionic numbers the associator~(\ref{ass}) is in general 
non-vanishing, however, the ``alternative condition'' is fulfilled
\bel{rul}
\{ x, \; y, \; z\}+\{ z, \; y, \; x\}=0 \q .
\ee

In 1989, writing a quaternionic Dirac equation~\cite{dir2}, Rotelli introduced
a {\tt barred}  momentum operator
\be
-\bfm{\dv}\mid i \qq [~(-\bfm{\dv}\mid i)\psi\equiv -\bfm{\dv}\psi i~] \q .
\ee
In a recent paper~\cite{qua2}, based upon the Rotelli operators, 
{\tt partially barred quaternions} 
\be
q+p\mid i \qq [~q, \; p \in \qu~] \q ,
\ee
have been used to formulate a quaternionic quantum mechanics. 

A complete generalization for quaternionic numbers is represented by 
the following barred operators
\be
q_{1} + q_{2}\mid i + q_{3}\mid j + q_{4}\mid k \qq
[~q_{1,...,4} \in \qu~] \q ,
\ee
which we call {\tt fully barred quaternions}, or simply barred  
quaternions. They, with their 16 linearly independent elements, form
a basis of $GL(4, \rea )$ and are successfully used 
to reformulate Lorentz space-time transformations~\cite{rel} and write down a 
one-component Dirac equation~\cite{dir4}.

Thus, it seems to us natural to investigate the existence of
{\tt barred octonions}
\be
{\cal O}_{0}+ \sum_{m=1}^{7} {\cal O}_{m}\mid e_{m} \qq 
[~ {\cal O}_{0,...,7} ~~ \mbox{octonions}~] \q .
\ee
Nevertheless, we must observe that an octonionic {\tt barred} operator, 
\bfm{a\mid b}, which acts on octonionic wave functions, $\psi$, 
\[ [~a\mid b~]~\psi \equiv a\psi b \q , \]
is not a well defined object. For $a\neq b$ the triple product $a\psi b$ 
could be either $(a\psi)b$ or $a(\psi b)$. So, in order to avoid the 
ambiguity due to the nonassociativity of 
the octonionic numbers, we need to define left/right-barred 
operators. We will indicate {\tt left-barred} operators by 
\bfm{a~)~b}, with $a$ and $b$ which represent octonionic numbers. They 
act on octonionic functions $\psi$ as follows
\bml
\be
[~a~)~b~]~\psi = (a\psi)b \q .
\ee
In similar way we can introduce {\tt right-barred} operators, defined by 
\bfm{a~(~b} ,
\be
[~a~(~b~]~\psi = a(\psi b) \q .
\ee
\eml
Obviously, there are barred-operators in which the nonassociativity is not 
of relevance, like 
\[ 1~)~a = 1~(~a \equiv 1\mid a \q . \]
Furthermore, from eq.~(\ref{rul}), we have
\[ \{ x, \; y, \; x\}=0 \q ,\]
so  
\[ a~)~a = a~(~a  \equiv a\mid a \q .\]
Besides, it is possible to prove, by eq.~(\ref{rul}), 
that each right-barred operator can be 
expressed by a suitable combination of left-barred operators. For further 
details, the reader can consult the mathematical paper~\cite{jmp}. So we can 
represent the most general octonionic operator by only 64 left-barred 
objects 
\bel{go}
{\cal O}_{0}+\sum_{m=1}^{7} {\cal O}_{m}~)~e_{m} \qq 
[~{\cal O}_{0, ...,7}~~ \mbox{octonions}~] \q .
\ee
This suggests a correspondence between our barred   
octonions and 
$GL(8, \rea)$ (a complete discussion about the 
above-mentioned relationship is given in the following section).

\section{Translation Rules}

In order to explain the idea of translation, let us look 
explicitly at the action of the operators $1\mid e_1$ and $e_2$,  on a generic 
octonionic  function $\vpg$
\be
\vpg = \vpg_0
 + e_1 \vpg_1 + e_2 \vpg_2 + e_3 \vpg_3
+ e_4 \vpg_4 + e_5 \vpg_5 + e_6 \vpg_6 + e_7 \vpg_7
\q [~\vpg_{0,\dots ,7} \in \rea~] \q .
\ee
We have
\bml
\beal{opa}
[~1\mid e_{1}~]~\vpg ~ \equiv ~\vpg e_1 & ~=~ & 
e_1 \vpg_0 - \vpg_1 - e_3 \vpg_2 + e_2 \vpg_3
- e_5 \vpg_4 + e_4 \vpg_5 + e_7 \vpg_6 - e_6 \vpg_7 \q , \\
 e_{2}\vpg & ~=~ & e_2 \vpg_0 - e_3 \vpg_1 - \vpg_2 + e_1 \vpg_3
+ e_6 \vpg_4 + e_7 \vpg_5 - e_4 \vpg_6 - e_5 \vpg_7
\q .
\eea
\eml
If we represent our octonionic function $\vpg$ by the following real column 
vector
\be
\vpg ~ \lrw ~ \left( \begin{array}{c}
\vpg_0\\
\vpg_1\\
\vpg_2\\
\vpg_3\\
\vpg_4\\
\vpg_5\\
\vpg_6\\
\vpg_7
\end{array}
\right) \q ,
\ee
we can rewrite the eqs.~(\ref{opa}-b) in matrix form,
\bml
\bea
\left(
\begin{array}{cccccccc}
0 & $-1$ & 0 & 0 & 0 & 0 & 0 &0\\
1 & 0 & 0 & 0 & 0 & 0 & 0 &0\\
0 & 0 & 0 & 1 & 0 & 0 & 0 &0\\
0 & 0 & $-1$& 0 & 0 & 0 & 0 &0\\
0 & 0 & 0 & 0 & 0 & 1 & 0 &0\\
0 & 0 & 0 & 0 &$ -1$& 0 & 0 &0\\
0 & 0 & 0 & 0 & 0 & 0 & 0 &$-1$\\
0 & 0 & 0 & 0 & 0 & 0 & 1 &0
\end{array}
\right)
\left( \begin{array}{c}
\vpg_0\\
\vpg_1\\
\vpg_2\\
\vpg_3\\
\vpg_4\\
\vpg_5\\
\vpg_6\\
\vpg_7
\end{array}
\right) & = &
\left( \begin{array}{c}
$-$\vpg_1\\
\vpg_0\\
\vpg_3\\
$-$\vpg_2\\
\vpg_5\\
$-$\vpg_4\\
$-$\vpg_7\\
\vpg_6
\end{array} \right) \q , \\
\left(
\begin{array}{cccccccc}
0 & 0 &$-1$ & 0 & 0 & 0 & 0 &0\\
0 & 0 & 0 & 1 & 0 & 0 & 0 &0\\
1 & 0 & 0 & 0 & 0 & 0 & 0 &0\\
0 & $-1$& 0 & 0 & 0 & 0 & 0 &0\\
0 & 0 & 0 & 0 & 0 & 0 & $-1$&0\\
0 & 0 & 0 & 0 & 0 & 0 & 0 &$-1$\\
0 & 0 & 0 & 0 & 1 & 0 & 0 &0\\
0 & 0 & 0 & 0 & 0 & 1 & 0 &0
\end{array}
\right)
\left( \begin{array}{c}
\vpg_0\\
\vpg_1\\
\vpg_2\\
\vpg_3\\
\vpg_4\\
\vpg_5\\
\vpg_6\\
\vpg_7
\end{array}
\right) & = &
\left( \begin{array}{c}
$-$\vpg_2\\
\vpg_3\\
\vpg_0\\
$-$\vpg_1\\
$-$\vpg_6 \\
$-$\vpg_7\\
\vpg_4\\
\vpg_5
\end{array} \right) \q . 
\eea
\eml
In this way we can immediately obtain a real matrix representation for the 
octonionic barred operators $1\mid e_{1}$ and $e_{2}$. Following this 
procedure we can construct the complete set of translation rules~\cite{jmp}. 

Let us now discuss of the relation between octonions and complex matrices. 
Complex groups play a critical role in physics. No one can deny the importance
of $U(1, \co)$ or $SU(2, \co)$. In relativistic
quantum mechanics, $\glc$ is essential in writing the 
Dirac equation. Having $\glr$, we should be able
to extract its subgroup $\glc$. So, we can 
translate the famous Dirac-gamma matrices and 
write down a new octonionic Dirac equation.

If we analyse the action of left-barred operators on our octonionic wave 
functions
\be
\psi = \psi_{1} + e_{2} \psi_{2} + e_{4} \psi_{3} + e_{6} \psi_{4} \qq
[~\psi_{1,  ..., 4} \in \bfm{\cal C}(1, \; e_{1})~] \q ,
\ee
we find, for example,
\bean
 & e_{2}\psi & ~=~  -\psi_{2} + e_{2} \psi_{1} -
e_{4} \psi_{4}^{*} + e_{6} \psi_{3}^{*} \q ,\\
~[~e_{3}~)~e_{1}~]~\psi ~\equiv ~ & (e_{3}\psi) e_{1} & ~=~ 
\psi_{2} + e_{2} \psi_{1} + e_{4} \psi_{4}^{*} - e_{6} \psi_{3}^{*} \q .
\eean
Obviously, the previous operators $e_2$ or $e_3~)~e_1$ cannot be 
represented by matrices, nevertheless we note that their  combined action 
gives us
\[
e_{2}\psi + (e_{3}\psi)e_{1} = 2 e_{2}\psi_{1} \q ,
\]
and it allows us to represent the octonionic barred operator 
\bml
\be 
e_{2} \; + \; e_{3}~)~e_{1} \q , 
\ee
by the $4\times 4$ complex matrix
\be
\bamq{0}{0}{0}{0} 2 & 0 & 0 & 0\\ 0 & 0 & 0 & 0\\ 0 & 0 & 0 & 0 \eam \q .
\ee
\eml
Following this procedure we can represent a generic $4\times 4$ complex 
matrix by octonionic barred operators. 
In Appendix B
we give the full basis of $\glc$ in terms of octonionic 
left-barred operators. It is clear that, only, particular 
combinations of left-barred operators is allowed to 
reproduce the associative matrix algebra. In order to make 
our discussion smooth, we refer the interested reader to the 
mathematical paper~\cite{jmp}. 
We can quickly relate 
$1\mid e_1$ with the complex matrix $i\openone_{4 \times 4}$ which will be 
relevant to an {\tt appropriate} definition for the octonionic momentum 
operator~\cite{oqm}. The operator $1\mid e_{1}$ (represented by the matrix 
$i \openone_{4\times 4}$) commutes with all operators which can be translated 
by  $4\times 4$ complex matrices. This is not generally true 
for a generic octonionic operator.  For example, we can show that the 
operator $1\mid e_{1}$ doesn't commute with $e_{2}$, explicitly
\bml
\bea
e_2 ~ \{ ~[~1\mid e_1 ~] ~\psi ~\} &~\equiv  e_{2}(\psi e_{1}) & ~=~ -e_1 \psi_2 - e_3 \psi_1 - 
e_5 \psi_4^* - e_7 \psi_3^* \q ,\\
~[~1\mid e_1 ~] ~ \{ e_2 ~\psi ~\} & ~\equiv  (e_{2} \psi) e_{1} & ~=~ -e_1 \psi_2 - e_3 \psi_1 + 
e_5 \psi_4^* + e_7 \psi_3^*
  \q .
\eea
\eml
The interpretation is simple: $e_{2}$ cannot be represented by a $4\times 4$ 
complex matrix.

We conclude this section by showing explicitly an octonionic representation 
for the Dirac gamma-matrices~\cite{itz}:\\
\new
\bc
{\tt Dirac representation,}
\bml
\beal{odgm1}
\cg^{0} & = & \frac{1}{3} -\frac{2}{3} \sum_{m=1}^{3} e_{m}\mid e_{m} +
\frac{1}{3} \sum_{n=4}^{7} e_{n} \mid e_{n} \q ,\\
\cg^{1} & = & -\frac{2}{3} e_{6} -\frac{1}{3}\mid e_{6} + e_{5}~)~e_{3} - 
e_{3}~)~e_{5}  - \frac{1}{3} \sum_{p, \; s =1}^{7} \eg_{ps6} e_{p}~)~e_{s} 
\q ,\\
\cg^{2} & = & -\frac{2}{3} e_{7} -\frac{1}{3}\mid e_{7} + e_{3}~)~e_{4} - 
e_{4}~)~e_{3}  - \frac{1}{3} \sum_{p, \; s =1}^{7} \eg_{ps7} e_{p}~)~e_{s} 
\q ,\\
\cg^{3} & = & -\frac{2}{3} e_{4} -\frac{1}{3}\mid e_{4} + e_{7}~)~e_{3} - 
e_{3}~)~e_{7}  - \frac{1}{3} \sum_{p, \; s =1}^{7} \eg_{ps4} e_{p}~)~e_{s} 
\q ;
\eea
\eml
\ec

\section{Octonionic Dirac Equation}

In the previous section we have given the gamma-matrices in three different 
octonionic representations. Obviously, we can investigate the 
possibility of having a more simpler representation for our octonionic 
$\cg^{\mu}$-matrices, without translation. 

Why not
\[ e_{1} \; , \q e_{2} \; , \q e_{3} \q \mbox{and} \q e_{4}\mid e_{4} ~~~~\]
or
\[ e_{1} \; , \q e_{2} \; , \q e_{3} \q \mbox{and} \q e_{4}~)~e_{1} \q ?\]
Apparently, they represent suitable choices. Nevertheless, the octonionic 
world is full of hidden traps and so we must proceed with prudence. 
Let us start from the standard Dirac equation
\be 
\cg^\nu p_{\nu} \psi=m\psi \q ,
\ee
(we discuss the momentum operator in the paper of ref.~\cite{oqm}, 
here, $p_{\nu}$ represents the ``real'' eigenvalue of the 
momentum operator) and apply 
$\cg^{\mu} p_{\mu}$ to our equation 
\be 
\cg^{\mu} p_{\mu}(\cg^{\nu} p_{\nu} \psi)=m \cg^{\mu} p_{\mu} \psi \q .
\ee
The previous equation can be concisely rewritten as
\be 
p^{\mu} p_{\nu} \cg^{\mu} (\cg^{\nu} \psi)=m^{2} \psi \q .
\ee
Requiring that each component of $\psi$ satisfy the standard Klein-Gordon 
equation  we find the Dirac condition, which becomes in the octonionic world
\bel{odc}
\cg^{\mu}(\cg^{\nu}\psi)+\cg^{\nu}(\cg^{\mu}\psi)=2g^{\mu \nu} \psi \q ,
\ee
(where the parenthesis are relevant because of the octonions nonassociative 
nature). Using octonionic numbers and no barred operators we can obtain, 
from~(\ref{odc}), the standard Dirac condition
\bel{sdc}
\{ \cg^{\mu}, \; \cg^{\nu} \} = 2 g^{\mu \nu} \q .
\ee
In fact, recalling the associator property 
[which follows from eq.~(\ref{eqass})] 
\[ \{a, \; b, \; \psi \} = - \{b, \; a, \; \psi \} \qq 
[~a, \; b \q \mbox{octonionic numbers}~] \q , \]
we quickly find the following correspondence relation 
\[ (ab+ba)\psi=a(b\psi)+b(a\psi) \q . \]
We have no problem to write down three suitable gamma-matrices which 
satisfy the Dirac condition~(\ref{sdc}),
\be
(\cg^{1}, \; \cg^{2}, \; \cg^{3}) \equiv (e_{1}, \; e_{2}, \; e_{3}) \q ,
\ee
but, barred operators like 
\[ e_{4}\mid e_{4} \q \mbox{or} \q e_{4}~)~e_{1} \]
cannot represent the matrix $\cg^{0}$. After straightforward algebraic 
manipulations, one can prove that 
the barred operator, $e_{4}\mid e_{4}$,  doesn't anticommute 
with $e_{1}$, 
\bea 
e_{1}(e_{4}\psi e_{4})+e_{4}(e_{1}\psi)e_{4} & = & 
-2 (e_{3}\psi_2 + e_7 \psi_4 ) \neq  0 \qq 
[~\psi=\psi_1 + e_2 \psi_2 +e_4 \psi_3 + e_6 \psi_4 ~] \q ,
\eea
whereas $e_{4}~)~e_{1}$ anticommutes with $e_1$
\bml
\bea 
e_{1}[(e_{4}\psi) e_{1}]+[e_{4}(e_{1}\psi)]e_{1} & = &  0 \q ,
\eea
but we know that $\gamma_0^2=1$, whereas 
\bea
\{ e_{4}[(e_{4}\psi) e_{1} ] \} e_1  & = & \psi_1 -e_2 \psi_2 +e_4 \psi_3
 -e_6 \psi_4 \neq \psi \q .
\eea
\eml

Thus, we must be satisfied with the octonionic representations given in the 
previous section. 

We recall that the appropriate momentum operator in OQM  with complex 
geometry~\cite{oqm} is
\[
{\cal P}^{\mu} \equiv \dv^{\mu} \mid e_{1} \q .
\]
Thus, the octonionic Dirac equation, in covariant form, is given by
\bel{ode}
\cg^{\mu}(\dv_{\mu}\psi e_{1})=m\psi \q ,
\ee
where $\cg^{\mu}$ are represented by octonionic barred  
operators~(\ref{odgm1}-d). We can now proceed in the standard manner. 
Plane wave solutions exist [${\bf p}~(\equiv -\bfm{\dv} \mid e_{1}$)  
commutes with a generic octonionic Hamiltonian] and are of the form
\be
\psi({\bf x}, \; t) = [~u_{1}({\bf p})+e_{2}u_{2}({\bf p})+
e_{4}u_{3}({\bf p})+e_{6}u_{4}({\bf p})~]~e^{-pxe_{1}} \qq
[~u_{1, ... , 4} \in \bfm{\cal C}(1, \; e_{1})~] \q .
\ee
Let's start with 
\[{\bf p} \equiv (0, \; 0, \; p_{z}) \q , \]
from~(\ref{ode}), we have
\bel{ode2}
E(\cg^{0} \psi) - p_{z}(\cg^{3} \psi) =m\psi \q .
\ee
Using the explicit form of the octonionic operators $\cg^{0, \; 3}$ and 
extracting their action (see appendix A) we find  
\bel{appc}
E(u_{1}+e_{2}u_{2}-e_{4}u_{3}-e_{6}u_{4})
-p_{z}(u_{3}-e_{2}u_{4}-e_{4}u_{1}+e_{6}u_{2})
=m(u_{1}+e_{2}u_{2}+e_{4}u_{3}+e_{6}u_{4}) 
\ee
From~(\ref{appc}), we derive four complex equations:
\bean
(E-m)u_{1} & = & +p_{z}u_{3} \q ,\\
(E-m)u_{2} & = & -p_{z}u_{4} \q ,\\
(E+m)u_{3} & = & +p_{z}u_{1} \q ,\\
(E+m)u_{4} & = & -p_{z}u_{2} \q .
\eean
After simple algebraic manipulations we find the following octonionic 
Dirac solutions:
\bean
E=+\vert E \vert & ~~~~~~~~u^{(1)}=N \left( 1 + e_{4} 
         \frac{p_{z}}{\vert E \vert +m} \right) \q , 
\q u^{(2)}= N \left( e_{2}-e_{6} \frac{p_{z}}{\vert E \vert +m} \right)
                                =u^{(1)}e_{2} \q ; \\
E=-\vert E \vert & ~~~~~~~~u^{(3)}=N \left(\frac{p_{z}}{\vert E \vert +m} 
- e_{4} \right) \q , 
\q  u^{(4)}=N \left(e_{2}\frac{p_{z}}{\vert E \vert +m} +e_{6} \right) 
                                =u^{(3)}e_{2} \q ,
\eean
with $N$ real normalization constant. Setting the norm to 
$2\vert E \vert$, we find
\[ N=(\vert E \vert + m)^{\frac{1}{2}} \q . \]

We now observe (as for the quaternionic Dirac equation) a difference with 
respect to the standard Dirac equation. 
Working in our representation~(\ref{odgm1}-d) and introducing the 
octonionic spinor
\[ \bar{u}\equiv(\cg_{0} u)^{+}= u_{1}^{*}-e_{2}u_{2}+
e_{4}u_{3}+e_{6}u_{4} \qq [~u= u_{1}+e_{2}u_{2}+
e_{4}u_{3}+e_{6}u_{4}~] \q ,\]
we have
\be
\bar{u}^{(1)}u^{(1)}=u^{(1)}\bar{u}^{(1)}=
\bar{u}^{(2)}u^{(2)}=u^{(2)}\bar{u}^{(2)}=2(m+e_{4}p_{z}) \q .
\ee
Thus we find 
\bml
\bel{os}
u^{(1)}\bar{u}^{(1)}+u^{(2)}\bar{u}^{(2)}=4(m+e_{4}p_{z}) \q ,
\ee
instead of the expected relation
\bel{cs}
u^{(1)}\bar{u}^{(1)}+u^{(2)}\bar{u}^{(2)}=\cg^{0} E - \cg^{3} p_{z} + m \q .
\ee
\eml
Furthermore, the previous difference is compensated if we compare the 
complex projection of~(\ref{os}) with the trace of~(\ref{cs})
\be
[~(u^{(1)}\bar{u}^{(1)}+u^{(2)}\bar{u}^{(2)})^{OQM}~]_{c}~\equiv~
Tr~[~(u^{(1)}\bar{u}^{(1)}+u^{(2)}\bar{u}^{(2)})^{CQM}~]~=~4m \q ,
\ee
which suggest to redefine the trace as ``complex'' trace. 
We know that spinor relations like~(\ref{os}-b) are relevant in 
perturbation calculus, so the previous results suggest to analyze quantum 
electrodynamics in order to investigate possible differences between 
complex and octonionic quantum field. This could represent the aim of 
a future work.

\con

In the physical literature, we find a method to partially overcome the issues 
relating to the octonions nonassociativity. Some people introduces a ``new'' 
imaginary units ``~$i=\sqrt{-1}$~'' which commutes with all others 
octonionic imaginary units, $e_{m}$. The new field is often called 
{\tt complexified octonionic field}. Different papers have been written in 
such a formalism: Quark Structure and Octonions~\cite{gur2}, 
Octonions, Quark and QCD~\cite{mor}, Dirac-Clifford algebra~\cite{edm}, 
Octonions and Isospin~\cite{pen}, and so on. 
In literature we also find a Dirac equation formulation by {\tt complexified} 
octonions with an embarrassing doubling of solutions: {\sl ``... the wave 
functions $\tilde{\psi}$ is not a column matrix, but must be taken as an 
octonion. $\tilde{\psi}$ therefore consists of eight wave functions, rather 
than the four wave functions of the Dirac equation''}~\cite{pen}. 
In this paper we have presented an alternative 
way to look at the octonionic world. 
No new imaginary unit is necessary to formulate in a consistent way an 
octonionic quantum mechanics.

Nevertheless complexified ring division algebras have been used in 
interesting works of Morita~\cite{mor2} to formulate the whole standard 
model.

Having a nonassociative algebra needs special care. In this work, we 
introduced a ``trick'' which allowed us to manipulate octonions without useless 
efforts. We summarize the more important results found in previous sections:

\bc
{\tt P - Physical Contents :}
\ec

{\tt P1} - We emphasize that a characteristic of our formalism is the 
{\em absolute need of a complex scalar product} (in QQM the use of a 
complex geometry is not obligatory and thus a question of choice). 
Using a complex geometry we 
overcame the hermiticity problem and gave the appropriate and unique 
definition of momentum operator;

{\tt P2} - A positive feature of this octonionic version of quantum 
mechanics, is the appearance of all four standard Dirac free-particle 
solutions notwithstanding the one-component structure of the wave functions. 
We have the following situation for the division algebras:
\bc
\bt{lcccl}
{\sf field} :~~~ & ~~complex,~~ & ~~quaternions,~~ & ~~octonions,~~& \\
{\sf Dirac Equation} :~~~ & $4\times 4$, &  $2\times 2$, &  $1\times 1$ & 
~~~{\fs ( matrix dimension ) ;} 
\et
\ec

{\tt P3} - Many physical result can be reobtained by translation, so we 
have one version of octonionic quantum mechanics where the standard 
quantum mechanics could be reproduced. This represents for the authors a first 
fundamental step towards an octonionic world. We remark that 
our translation will not be possible in 
all situations, so it is only partial, consistent with the fact that the 
octonionic version could provide additional physical predictions. 

\bc
{\tt I - Further Investigations :}
\ec

We list some open questions for future investigations, 
whose study lead to further insights.

{\tt I4} - The reproduction in octonionic calculations of the standard QED 
results will be a nontrivial objective, due to the explicit differences in 
certain spinorial identities (see section IV). We are going to study 
this problem in a forthcoming paper;

{\tt I5} - A very attractive point is to try to treat the strong field 
by octonions, and then to formulate in a suitable manner a standard 
model, based on our octonionic dynamical Dirac equation.

We conclude emphasizing that the core of our paper is surely represented by 
absolute need of adopting a complex geometry within a quantum octonionic 
world.

\section*{Appendix A\\
$\cg^{0, \; 3}$-action on octonionic spinors}

In the following tables,  we explicitly show the action on the octonionic 
spinor 
\[
u=u_{1}+e_{2}u_{2}+e_{4}u_{3}+e_{6}u_{4} \qq [~u_{1,...,4} \in 
\bfm{\cal C}(1, \; e_{1})~] \q , \]
of the barred operators which appear in $\cg^{0}$ and $\cg^{3}$. Using such 
tables, after straightforwards algebraic manipulations we find
\bean
\cg^{0}u & ~=~ & u_{1}+e_{2}u_{2}-e_{4}u_{3}-e_{6}u_{4} \q ,\\
\cg^{3}u & ~=~ & u_{3}-e_{2}u_{4}-e_{4}u_{1}+e_{6}u_{2} \q . 
\eean
\vs{.5cm}
\bc
\bt{l|rrrr}
 & & & & \\
$\cg^{0}$-action~~~ & 
$~~~~~~~u_{1}$ & $~~~~~~~e_{2}u_{2}$ & $~~~~~~~e_{4}u_{3}$ & 
$~~~~~~~e_{6}u_{4}$\\
 & & & & \\
\hline \hline
 & & & & \\
$e_{1}\mid e_{1}$ &
$-u_{1}$ & $e_{2}u_{2}$ & $e_{4}u_{3}$ & $e_{6}u_{4}$\\
$e_{2}\mid e_{2}$ &
$-u_{1}^{*}$ & $-e_{2}u_{2}^{*}$ & $e_{4}u_{3}$ & $e_{6}u_{4}$\\
$e_{3}\mid e_{3}$ &
$-u_{1}^{*}$ & $e_{2}u_{2}^{*}$ & $e_{4}u_{3}$ & $e_{6}u_{4}$\\
$e_{4}\mid e_{4}$ &
$-u_{1}^{*}$ & $e_{2}u_{2}^{*}$ & $-e_{4}u_{3}^{*}$ & $e_{6}u_{4}$\\
$e_{5}\mid e_{5}$ &
$-u_{1}^{*}$ & $e_{2}u_{2}$ & $e_{4}u_{3}^{*}$ & $e_{6}u_{4}$\\
$e_{6}\mid e_{6}$ &
$-u_{1}^{*}$ & $e_{2}u_{2}$ & $e_{4}u_{3}$ & $-e_{6}u_{4}^{*}$\\
$e_{7}\mid e_{7}$ &
$-u_{1}^{*}$ & $e_{2}u_{2}$ & $e_{4}u_{3}$ & $e_{6}u_{4}^{*}$
\et
\ec
\vs{.5cm}
\bc
\bt{l|rrrr}
 & & & & \\
$\cg^{3}$-action~~~ & 
$~~~~~~~u_{1}$ & $~~~~~~~e_{2}u_{2}$ & $~~~~~~~e_{4}u_{3}$ & 
$~~~~~~~e_{6}u_{4}$\\
 & & & & \\
\hline \hline 
 & & & & \\
$e_{4}$ &
$e_{4}u_{1}$ &  $-e_{6}u_{2}^{*}$ & $-u_{3}$ & $e_{2}u_{4}$\\
$1\mid e_{4}$ &
$e_{4}u_{1}^{*}$ &  $e_{6}u_{2}^{*}$ & $-u_{3}^{*}$ & $-e_{2}u_{4}^{*}$\\
$e_{7}~)~e_{3}$ &
$e_{4}u_{1}^{*}$ &  $e_{6}u_{2}$ & 
$u_{3}$ & $-e_{2}u_{4}^{*}$\\
$e_{3}~)~e_{7}$ &
$-e_{4}u_{1}^{*}$ &  $-e_{6}u_{2}^{*}$ & 
$-u_{3}$ & $e_{2}u_{4}$\\
$e_{6}~)~e_{2}$ &
$e_{4}u_{1}^{*}$ &  $-e_{6}u_{2}$ & 
$u_{3}$ & $-e_{2}u_{4}^{*}$\\
$e_{2}~)~e_{6}$ &
$-e_{4}u_{1}^{*}$ &  $-e_{6}u_{2}^{*}$ & 
$-u_{3}$ & $-e_{2}u_{4}$\\
$e_{5}~)~e_{1}$ &
$e_{4}u_{1}$ &  $e_{6}u_{2}^{*}$ & $u_{3}$ & 
$-e_{2}u_{4}^{*}$\\
$e_{1}~)~e_{5}$ &
$-e_{4}u_{1}^{*}$ &  $-e_{6}u_{2}^{*}$ & $-u_{3}^{*}$ & 
$e_{2}u_{4}^{*}$
\et
\ec

\new

\section*{Appendix B}

In the following charts we establish the connection between $4\times 4$ 
complex matrices and octonionic left/right-barred operators. We indicate with 
${\cal R}_{mn}$ (${\cal C}_{mn}$)  the $4\times 4$ real (complex) matrices 
with 1 ($i$) in $mn$-element and zeros elsewhere.\\ 
\bc
{\tt $4 \times 4$ complex matrices and left-barred operators:}
\ec
\bean
{\cal R}_{11} & ~\lrw~ & \frac{1}{2}~[~1-e_{1}\mid e_{1}~] \\ 
{\cal R}_{12} & ~\lrw~ & \frac{1}{6}~[~2 e_{1}~)~e_{3} + e_{3}~)~e_{1} - 
2 \mid e_{2} - e_{2} + e_{4}~)~e_{6} - e_{6}~)~e_{4} + 
e_{5}~)~e_{7} - e_{7}~)~e_{5} ~] \\ 
{\cal R}_{13} & ~\lrw~ & \frac{1}{6}~[~2 e_{1}~)~e_{5} + e_{5}~)~e_{1} - 
2 \mid e_{4} - e_{4} + e_{6}~)~e_{2} - e_{2}~)~e_{6} + 
e_{7}~)~e_{3} - e_{3}~)~e_{7} ~] \\ 
{\cal R}_{14} & ~\lrw~ & \frac{1}{6}~[~2 e_{1}~)~e_{7} + e_{7}~)~e_{1} - 
2 \mid e_{6} - e_{6} + e_{2}~)~e_{4} - e_{4}~)~e_{2} + 
e_{5}~)~e_{3} - e_{3}~)~e_{5} ~] \\ 
{\cal R}_{21} & ~\lrw~ & \frac{1}{2}~[~e_{2} + e_{3}~)~e_{1} ~] \\ 
{\cal R}_{22} & ~\lrw~ & \frac{1}{6}~[~1+e_{1}\mid e_{1}+e_{4}\mid e_{4}+
e_{5}\mid e_{5}+e_{6}\mid e_{6}+e_{7}\mid e_{7}~] -
\frac{1}{3}~[~e_{2}\mid e_{2}+e_{3}\mid e_{3}~]\\
{\cal R}_{23} & ~\lrw~ & \frac{1}{2}~[~-e_{2}~)~e_{4} - e_{3}~)~e_{5} ~] \\ 
{\cal R}_{24} & ~\lrw~ & \frac{1}{2}~[~e_{3}~)~e_{7} - e_{2}~)~e_{6} ~] \\ 
{\cal R}_{31} & ~\lrw~ & \frac{1}{2}~[~e_{4} + e_{5}~)~e_{1} ~] \\ 
{\cal R}_{32} & ~\lrw~ & \frac{1}{2}~[~-e_{5}~)~e_{3} - e_{4}~)~e_{2} ~] \\ 
{\cal R}_{33} & ~\lrw~ & \frac{1}{6}~[~1+e_{1}\mid e_{1}+e_{2}\mid e_{2}+
e_{3}\mid e_{3}+e_{6}\mid e_{6}+e_{7}\mid e_{7}~] -
\frac{1}{3}~[~e_{4}\mid e_{4}+e_{5}\mid e_{5}~]\\
{\cal R}_{34} & ~\lrw~ & \frac{1}{2}~[~e_{5}~)~e_{7} - e_{4}~)~e_{6} ~] \\ 
{\cal R}_{41} & ~\lrw~ & \frac{1}{2}~[~e_{6} - e_{7}~)~e_{1} ~] \\ 
{\cal R}_{42} & ~\lrw~ & \frac{1}{2}~[~e_{7}~)~e_{3} - e_{6}~)~e_{2} ~] \\ 
{\cal R}_{43} & ~\lrw~ & \frac{1}{2}~[~e_{7}~)~e_{5} - e_{6}~)~e_{4} ~] \\ 
{\cal R}_{44} & ~\lrw~ & \frac{1}{6}~[~1+e_{1}\mid e_{1}+e_{2}\mid e_{2}+
e_{3}\mid e_{3}+e_{4}\mid e_{4}+e_{5}\mid e_{5}~] -
\frac{1}{3}~[~e_{6}\mid e_{6}+e_{7}\mid e_{7}~]\\
{\cal C}_{11} & ~\lrw~ & \frac{1}{2}~[~1\mid e_{1}+e_{1}~] \\ 
{\cal C}_{12} & ~\lrw~ & \frac{1}{6}~[~-2 e_{1}~)~e_{2} - e_{3} - 2 \mid e_{3}
- e_{2}~)~e_{1}   + e_{4}~)~e_{7} + e_{6}~)~e_{5} - e_{5}~)~e_{6} - 
e_{7}~)~e_{4}  ~] \\ 
{\cal C}_{13} & ~\lrw~ & \frac{1}{6}~[~-2 e_{1}~)~e_{4} - e_{5} - 2 \mid e_{5} 
- e_{4}~)~e_{1} - e_{6}~)~e_{3} - e_{2}~)~e_{7}  + e_{7}~)~e_{2} 
+ e_{3}~)~e_{6} ~] \\ 
{\cal C}_{14} & ~\lrw~ & \frac{1}{6}~[~-2 e_{1}~)~e_{6} + e_{7} + 2 \mid e_{7} 
- e_{6}~)~e_{1} - e_{2}~)~e_{5} + 
e_{4}~)~e_{3} + e_{5}~)~e_{2}  - e_{3}~)~e_{4} ~] \\ 
{\cal C}_{21} & ~\lrw~ & \frac{1}{2}~[~-e_{3} + e_{2}~)~e_{1} ~] \\ 
{\cal C}_{22} & ~\lrw ~ & \frac{1}{6}~[~1\mid e_{1} - e_{1} + e_{4}~)~e_{5}
-e_{5}~)~e_{4}-e_{6}~)~ e_{7}+e_{7}~)~ e_{6}~] -
\frac{1}{3}~[~e_{2}~)~ e_{3}-e_{3}~)~e_{2}~]\\
{\cal C}_{23} & ~\lrw~ & \frac{1}{2}~[~ - e_{2}~)~e_{5} + e_{3}~)~e_{4} ~] \\ 
{\cal C}_{24} & ~\lrw~ & \frac{1}{2}~[~ e_{3}~)~e_{6} + e_{2}~)~e_{7}~] \\ 
{\cal C}_{31} & ~\lrw~ & \frac{1}{2}~[~- e_{5} + e_{4}~)~e_{1} ~] \\ 
{\cal C}_{32} & ~\lrw~ & \frac{1}{2}~[~e_{5}~)~e_{2} - e_{4}~)~e_{3} ~] \\ 
{\cal C}_{33} & ~\lrw~ & \frac{1}{6}~[~1 \mid e_{1} - e_{1} +e_{2}~)~ e_{3}
-e_{3}~)~ e_{2}-e_{6}~)~ e_{7}+e_{7}~)~e_{6}~] -
\frac{1}{3}~[~e_{4}~)~ e_{5}-e_{5}~)~e_{4}~]\\
{\cal C}_{34} & ~\lrw~ & \frac{1}{2}~[~e_{5}~)~e_{6} + e_{4}~)~e_{7} ~] \\ 
{\cal C}_{41} & ~\lrw~ & \frac{1}{2}~[~ e_{7} + e_{6}~)~e_{1} ~] \\ 
{\cal C}_{42} & ~\lrw~ & \frac{1}{2}~[~- e_{7}~)~e_{2} - e_{6}~)~e_{3}  ~] \\ 
{\cal C}_{43} & ~\lrw~ & \frac{1}{2}~[~- e_{7}~)~e_{4} -e_{6}~)~e_{5}  ~] \\ 
{\cal C}_{44} & ~\lrw~ & \frac{1}{6}~[~1\mid e_{1} - e_{1} +e_{2}~)~e_{3}
-e_{3}~)~ e_{2}+e_{4}~)~ e_{5}-e_{5}~)~ e_{4}~] -
\frac{1}{3}~[~e_{7}~)~ e_{6}-e_{6}~)~ e_{7}~]
\eean
\\

\bref
\bi{gur1}
F.~G\"ursey, {\it Symmetries in Physics (1600-1980): Proc.~of the 1st 
International Meeting on the History of Scientific Ideas}, Seminari 
d'~Hist\`oria de les Ci\`ences, Barcelona, Spain, 1987, p.~557.
\bi{pais}
A.~Pais, \pxh{7}{291}{61}.
\bi{gur2}
M.~G\"unaydin and F.~G\"ursey, \jxe{14}{1651}{73}; \pxf{9}{3387}{74}.
\bi{mor}
K.~Morita, \pxxa{65}{787}{81}.
\bi{dix}
G.~Dixon, \nxd{B105}{349}{90}.
\bi{gur3}
F.~G\"ursey, {\it Yale Preprint C00-3075-178} (1978).
\bi{edm}
J.~D.~Edmonds, \pxa{5}{56}{92}.
\bi{jos1}
A.~Waldron and G.~C.~Joshi, {\it Melbourne Preprint UM-P-92/60} (1992).
\bi{jos2}
G.~C.~Lassig and G.~C.~Joshi, {\it Melbourne Preprint UM-P-95/09} (1995).\\
A.~Ritz and G.~C.~Joshi, {\it Melbourne Preprint UM-P-95/69} (1995).
\bi{dav}
A.~J.~Davies and G.~C.~Joshi, \jxe{27}{3036}{86}.
\bi{sup1}
T.~Kugo and P.~Townsend, \nxb{B221}{357}{87}.
\bi{sup2}
B.~Julia, {\it Lptens Preprint 82/14} (1982).
\bi{adl}
S.~L.~Adler, {\it Quaternionic Quantum Mechanics and Quantum Fields} 
(Oxford, New York, 1995).
\bi{adl1}
S.~L.~Adler,  \nxb{B415}{195}{94}.
\bi{qua1}
S.~De Leo and P.~Rotelli, \pxf{45}{575}{92}; \nxd{B110}{33}{95};\\
S.~De Leo, \pxxa{94}{11}{95}; {\it Quaternions for GUTs}, 
Int.~J.~Theor.~Phys. (submitted).
\bi{qua2}
S.~De Leo and P.~Rotelli, \pxxa{92}{917}{94}; {\it Odd Dimensional 
Translations between Complex and Quaternionic Quantum Mechanics} 
(to be published in Prog.~Theor.~Phys.).
\bi{qua3} 
S.~De Leo and P.~Rotelli, \ixa{10}{4359}{95}; 
Mod.~Phys.~Lett.~A {\bf 11}, 357 (1996).\\
S.~De Leo and P.~Rotelli, {\it Quaternionic Electroweak Theory}, 
J.~Phys.~G (submitted).
\bi{dir1}
S.~L.~Adler, \pxi{221B}{39}{89}.
\bi{dir2}
P.~Rotelli, \mxb{4}{933}{89}.
\bi{dir3}
A.~J.~Davies, \pxf{41}{2628}{90}.
\bi{dir4}
S.~De Leo, {\it One-component Dirac Equation}, Int.~J.~Mod.~Phys.~A 
(to be published).
\bi{hor}
L.~P.~Horwitz and L.~C.~Biedenharn, \axp{157}{432}{84}.\\ 
J.~Rembieli\'nski, \jxg{11}{2323}{78}.
\bi{rk}
K.~Abdel-Khalek and P.~Rotelli, {\em Quaternionic 
Supersymmetry}, in preparation.
\bi{jmp}
S.~De Leo and K.~Abdel-Khalek, {\em Octonionic Representations of 
$GL(8, {\cal R})$ and $GL(4, {\cal C})$}, J.~Math.~Phys. 
(submitted), hep-th/9607140. 
\bi{rel}
S.~De Leo, {\it Quaternions and Special Relativity}, J.~Math.~Phys. 
(to be published).
\bi{oqm}
S.~De Leo and K.~Abdel-Khalek, {\em Octonionic Quantum Mechanics and 
Complex Geometry}, Prog.~Theor.~Phys. (submitted).
\bi{itz}
The eqs.~(\ref{odgm1}-d) 
represent the octonionic 
counterpart of the complex matrices given on pag.~49 of the book:\\
C.~Itzykson and J.~B.~Zuber, {\it Quantum Field Theory} 
(McGraw-Hill, New York, 1985).
\bi{pen}
R.~Penney, \nxd{B3}{95}{71}.
\bi{mor2}
K.~Morita, \pxxa{67}{1860}{81}; \xxx{68}{2159}{82}; \xxx{70}{1648}{83}; 
\xxx{72}{1056}{84}; \xxx{73}{999}{84}; \xxx{75}{220}{85}; \xxx{90}{219}{93}. 

\eref

%%%%%%%%%%%%%%%%%%
\end{document}